\documentclass[11pt]{article}
\usepackage{amssymb,epsfig}
\begin{document}
\title{
\begin{flushright}
\normalsize
\begin{tabular}{r}
DFTT 25/98\\
UWThPh-1998-24\\
hep-ph/9805411
\end{tabular}
\end{flushright}
\vspace{1cm}
\textbf{Neutrino masses and mixing}\thanks{Talk presented by S.M. Bilenky at the
12$^{\mathrm{th}}$
\textit{Les Rencontres de Physique de la Vall\'ee d'Aoste:
Results and Perspectives in Particle Physics}
1-7 Mar 1998, La Thuile, Aosta Valley, Italy.}
}
\author{
S.M. Bilenky\\
\textit{Joint Institute for Nuclear Research, Dubna, Russia}\\[3mm]
C. Giunti\\
\textit{INFN, Sezione di Torino, and Dipartimento di Fisica Teorica,
Universit\`a di Torino,}\\
\textit{Via P. Giuria 1, I--10125 Torino, Italy}\\[3mm]
W. Grimus\\
\textit{Institute for Theoretical Physics, University of Vienna,}\\
\textit{Boltzmanngasse 5, A--1090 Vienna, Austria}
}
\date{May 1998}
\maketitle
\begin{abstract}
We present a short review
of the present status of the problem
of neutrino masses and mixing.
The existing experimental results indicate
that there are at least four massive neutrinos.
We show that
only two schemes with mixing of four neutrinos
and mass spectra in which two groups of close masses are separated
by the ``LSND gap''
($ \sim 1 \, \mathrm{eV} $)
are compatible with the results of all neutrino oscillation
experiments.
We discuss different consequences of these schemes
for future neutrino oscillation experiments.
\end{abstract}

\section{Introduction}
\label{Introduction}

The problem of neutrino masses and mixing
(see \cite{BP78,BP87,Mohapatra-Pal,CWKim})
is the most important problem of 
today's neutrino physics.
There are at present different indications
that neutrinos have small masses and that there is neutrino mixing.
These indications were obtained in
solar neutrino experiments
\cite{Homestake}--\cite{SK},
in atmospheric neutrino experiments
\cite{Kam-atm}--\cite{SK-atm},\cite{SK}
and in the LSND experiment
\cite{LSND}.
If the indications in favor of neutrino oscillations will be confirmed,
they will represent the first observation
of processes in which lepton numbers are not conserved.
It is generally 
believed that the investigation of such processes will allow us to
investigate
the physics at a scale much larger than the electroweak scale.

All the existing data on the investigation of the
\emph{weak interaction processes}
in which
neutrinos take part are perfectly described by the standard model of
electroweak interactions. There are two classes of electroweak 
interactions:
\begin{enumerate}

\item
Charged current (CC) interactions described by the Lagrangian
\begin{equation}
\mathcal{L}_{I}^{\mathrm{CC}}
=
- \frac{g}{2\sqrt{2}} \,
j^{\mathrm{CC}}_{\alpha} \, W^{\alpha}
+
\mathrm{h.c.}
\,,
\label{001}
\end{equation}
where
$g$ is the dimensionless SU(2) coupling constant
and the charged current
$j^{\mathrm{CC}}_{\alpha}$
is given by
\begin{equation}
j^{\mathrm{CC}}_{\alpha}
=
2
\sum_{\ell=e,\mu,\tau} \overline{\nu_{{\ell}L}} \, \gamma_{\alpha} \, \ell_{L}
+
\ldots
\,.
\label{002}
\end{equation}

\item
Neutral current (NC) interactions described by the Lagrangian
\begin{equation}
\mathcal{L}_{I}^{\mathrm{NC}}
=
- \frac{g}{2\cos_{\mathrm{W}}} \,
j^{\mathrm{NC}}_{\alpha} \, Z^{\alpha}
\,,
\label{003}
\end{equation}
where
$\theta_{\mathrm{W}}$
is the Weinberg angle
and the neutral current
$j^{\mathrm{NC}}_{\alpha}$
is given by
\begin{equation}
j^{\mathrm{NC}}_{\alpha}
=
\sum_{\ell=e,\mu,\tau} \overline{\nu_{{\ell}L}} \, \gamma_{\alpha} \, \nu_{{\ell}L}
+
\ldots
\,.
\label{004}
\end{equation}

\end{enumerate}

Charged and neutral current
weak interactions conserve the total electron, muon and
tau lepton numbers
$L_e$, $L_\mu$, $L_\tau$
and the CC interactions
determine the \emph{notion of flavour neutrinos}
$\nu_e$, $\nu_\mu$, $\nu_\tau$.
For example,
we call muon neutrino $\nu_\mu$ the particle
that is produced in $\pi^{+}$-decay together with
a $\mu^{+}$
and so on.

The number of light flavour neutrinos
$n_\nu$
is equal to three.
This number was obtained in LEP experiments
from the measurement of the width of the decay
$ Z \to \nu + \bar\nu $.
The combined result of LEP experiments is
\cite{PDG}
\begin{equation}
n_\nu = 2.991 \pm 0.016
\,.
\label{005}
\end{equation}

The hypothesis of neutrino mixing,
initiated by B. Pontecorvo as early as 1957 \cite{Pontecorvo57},
is based on the assumption that neutrinos are massive particles and that 
the neutrino
mass term does not conserve lepton numbers.
After the standard procedure of 
the diagonalization of
the lepton-numbers non-conserving
neutrino mass term,
for the flavour neutrino field we have
\begin{equation}
\nu_{{\ell}L}
=
\sum_{i}
U_{{\ell}i} \, \nu_{iL}
\,,
\label{006}
\end{equation}
where
$\nu_{iL}$
is the field of neutrinos with mass $m_i$ and $U$ is
a unitary mixing matrix.

There are two possibilities for the fields of massive neutrinos:
\begin{enumerate}

\item
If the mass term conserves the total lepton number
\begin{equation}
L = L_e + L_\mu + L_\tau
\,,
\label{007}
\end{equation}
then
the fields $\nu_i$ are four-component Dirac fields
and the number of massive neutrinos is equal to the number
of flavour neutrinos, $n_\nu=3$.

Notice that a Dirac mass term can be generated by the standard Higgs 
mechanism by adding right-handed neutrino gauge singlets in the same
way as the mass terms of all the other fundamental fermions.
In this case,
the numerous parameters of the Standard Model
will be increased by the addition of the
neutrino masses and mixing angles.
In the framework of the Standard Model there is no mechanism that can 
explain the smallness of neutrino masses.

\item
If the conservation of the total lepton number is violated,
the fields of neutrinos 
with definite masses are Majorana fields,
i.e. fields of particles
with all charges equal to zero.
The Majorana neutrino fields $\nu_i$ satisfy
the condition 
\begin{equation}
\nu_i^c = \mathcal{C} \, \overline{\nu}_i^T = \nu_i
\,,
\label{008}
\end{equation}
where $\mathcal{C}$ is the matrix of charge conjugation
which is determined by
\begin{equation}
\mathcal{C}
\,
\gamma_{\alpha}^{T}
\,
\mathcal{C}^{-1}
=
- \gamma_{\alpha}
\,,
\quad
\mathcal{C}^{\dagger} = \mathcal{C}^{-1}
\,,
\quad
\mathcal{C}^{T} = - \mathcal{C}
\,.
\label{009}
\end{equation}
A Majorana mass term can be generated only in the framework
of models beyond the Standard Model
(see \cite{Mohapatra-Pal}). 

The number $n$ of massive Majorana neutrinos
can be equal or larger than
the number of flavour neutrinos ($n_\nu=3$).
If $n>3$,
for the mixing we have
\begin{equation}
\nu_{{\ell}L}
=
\sum_{i=1}^{n}
U_{{\ell}i} \, \nu_{iL}
\,,
\quad
(\nu_{aR})^c
=
\sum_{i=1}^{n}
U_{ai} \, \nu_{iL}
\,.
\label{010}
\end{equation} 
where $U$ is a $n{\times}n$ unitary mixing matrix and  
the fields
$\nu_{aR}$
do not enter in the standard CC and NC
(the fields $\nu_{aR}$ are called sterile).
If $ m_i \ll m_Z $ for all $i=1,\ldots,n$,
the width of the decay
$ Z \to \nu + \bar\nu $
is determined only by the number of flavour neutrinos.
Let us stress, however, that because of the mixing (\ref{010}),
active neutrinos
$\nu_e$, $\nu_\mu$, $\nu_\tau$
can transform (in vacuum or in matter) 
into undetectable sterile states.

\end{enumerate}

From the existing data it follows that neutrino masses (if any) are much 
smaller than the masses of charged leptons and quarks.
The understanding of this 
phenomena is a big theoretical challenge.
A possible explanation of the smallness of 
neutrino masses is provided by the see-saw mechanism 
\cite{see-saw}.
This mechanism is based on the assumption that lepton numbers
are violated by the right-handed Majorana mass term at a scale
$M$ that is much larger than the scale
of electroweak symmetry breaking.
If the neutrino masses are of see-saw origin we have
the following consequences:
\begin{enumerate}

\item
Massive neutrinos are Majorana particles;

\item
The number of light massive neutrinos is equal to three;

\item
The neutrino masses are given by the see-saw formula
\begin{equation}
m_i \sim \frac{ ( m_i^F )^2 }{ M }
\ll
m_i^F
\quad
(i=1,2,3)
\,.
\label{011}
\end{equation}
where $m_i^F$ is the mass of
the charged lepton or up-quark in the $i^{\mathrm{th}}$ generation.
From Eq.(\ref{011}) it follows that in the see-saw case
the neutrino masses satisfy the hierarchy relation
\begin{equation}
m_1 \ll m_2 \ll m_3
\,.
\label{012}
\end{equation}

\end{enumerate}

We will finish this introduction with a brief review of
the experimental 
situation.
Indications in the favour of neutrino oscillations were found
in the following experiments:
\begin{enumerate}

\item
In all solar neutrino experiments:
Homestake \cite{Homestake},
Kamiokande \cite{Kam-sun},
GALLEX \cite{GALLEX},
SAGE \cite{SAGE}
and
Super-Kamiokande \cite{SK-sun,SK};

\item
In the
Kamiokande \cite{Kam-atm},
IMB \cite{IMB},
Soudan \cite{Soudan}
and
Super-Kamiokande \cite{SK-atm,SK}
atmospheric neutrino experiments;

\item
In the accelerator LSND experiment \cite{LSND}.

\end{enumerate}
From the analysis of the data
of these experiments it follows that there exist
\emph{three different scales}
of neutrino mass squared difference:
\begin{eqnarray}
&
\Delta{m}^2_{\mathrm{sun}}
\sim
10^{-5} \, \mathrm{eV}^2
\, (\mbox{MSW})
\quad \mbox{or} \quad
\Delta{m}^2_{\mathrm{sun}}
\sim
10^{-10} \, \mathrm{eV}^2
\, (\mbox{vac. osc.})
\quad
\mbox{\cite{HL97,FLM97}}
\,,
&
\label{SUNrange}
\\
&
\Delta{m}^2_{\mathrm{atm}}
\sim
5 \times 10^{-3} \, \mathrm{eV}^2
\quad
\mbox{\cite{Valencia}}
\,,
&
\label{ATMrange}
\\
&
\Delta{m}^2_{\mathrm{LSND}} \sim 1 \, \mathrm{eV}^2
\quad
\mbox{\cite{LSND}}
\,.
&
\label{LSNDrange}
\end{eqnarray}
The two possibilities for
$\Delta{m}^2_{\mathrm{sun}}$
correspond,
respectively,
to the
MSW \cite{MSW}
and
to the
vacuum oscillation
solutions of the solar neutrino problem.

On the other hand,
no indication in favour of neutrino oscillations was found
in numerous short-baseline (SBL) reactor and accelerator experiments
(see the review in Ref.\cite{Brunner}).
Also in the first long-baseline (LBL)
reactor experiment CHOOZ \cite{CHOOZ}
neutrinos oscillations were not found.

No indications in favour of non-zero neutrino masses were found
in the experiments on the measurement of the high-energy part of
the $\beta$-spectrum
in the decay
$
{}^3\mathrm{H}
\to
{}^3\mathrm{He}^+ + e^- + \bar\nu_e
$.
The following upper bounds for the effective neutrino mass
were found
in the Troitsk \cite{Troitsk} and Mainz \cite{Mainz}
experiments:
\begin{displaymath}
\begin{array}{lcccc} \displaystyle
\null & \null \displaystyle \null \quad \null & \null \displaystyle
m_{\nu}
\null & \null \displaystyle \null \quad \null & \null \displaystyle
m_{\nu}^2
\\ \displaystyle
\mbox{Troitsk}
\null & \null \displaystyle \null \quad \null & \null \displaystyle
< 3.9 \, \mathrm{eV}
\null & \null \displaystyle \null \quad \null & \null \displaystyle
1.5 \pm 5.9 \pm 3.6 \, \mathrm{eV}^2
\\ \displaystyle
\mbox{Mainz}
\null & \null \displaystyle \null \quad \null & \null \displaystyle
< 5.6 \, \mathrm{eV}
\null & \null \displaystyle \null \quad \null & \null \displaystyle
-22 \pm 17 \pm 14 \, \mathrm{eV}^2
\end{array}
\end{displaymath}

Many experiments on the search for neutrinoless double-beta decay
($(\beta\beta)_{0\nu}$), 
\begin{equation}
(A,Z)
\to
(A,Z+2)
+
e^{-}
+
e^{-}
\,,
\label{021}
\end{equation}
have been done.
This process is possible only if neutrinos are 
massive and Majorana particles.
The matrix element of the process is 
proportional to the effective Majorana mass
\begin{equation}
\langle{m}\rangle
=
\sum_{i}
U_{ei}^2
\,
m_{i}
\,.
\label{022}
\end{equation}
The $(\beta\beta)_{0\nu}$ process (\ref{021}) was not observed.
The Heidelberg-Moscow $(\beta\beta)_{0\nu}$ experiment \cite{Heidelberg-Moscow}
reached the following lower limit
for the half-live of $^{76}$Ge:
\begin{equation}
T_{1/2}(^{76}\mathrm{Ge})
>
1.2 \times 10^{25} \, \mathrm{y}
\qquad \mbox{(90\% CL)}
\,.
\label{023}
\end{equation}
From this result it follows that \cite{Heidelberg-Moscow}
\begin{equation}
|\langle{m}\rangle|
<
( 0.5 - 1.5 ) \, \mathrm{eV}
\,.
\label{024}
\end{equation}
Let us notice that in the next years the sensitivity of
$(\beta\beta)_{0\nu}$
experiments will reach
$ |\langle{m}\rangle| \simeq 0.1 \, \mathrm{eV}$
\cite{futureBB}.

In the analysis of the data of neutrino oscillation experiments it is important 
to take into account the data of all experiments
because different observables
are connected by the unitarity of the mixing matrix.
It is clear that this cannot be done in the usual framework
of two-neutrino mixing.
Thus,
the general case of $n$-neutrino mixing (see \cite{BP87})
must be considered.
We followed this approach in
Refs.\cite{BBGK95}--\cite{BGG98}.
We tried to answer to the following questions:
\begin{enumerate}

\item
Which neutrino mass spectrum is compatible with the data;

\item
What information on the elements of the neutrino mixing matrix
can be obtained from the data of SBL experiments;

\item
Which are the predictions for future LBL experiments.

\end{enumerate}

\section{Three massive neutrinos}
\label{Three massive neutrinos}

Let us consider first the case of three massive neutrinos
and a neutrino mass hierarchy
\cite{three,BBGK95,BBGK96,BGKM},
$ m_1 \ll m_2 \ll m_3 $.
We assume that
$ \Delta{m}^2_{21} \equiv m_2^2 - m_1^2 $
is relevant for the suppression of the flux
of solar neutrinos and
$ \Delta{m}^2 \equiv \Delta{m}^2_{31} \equiv m_3^2 - m_1^2 $
is relevant for the LSND anomaly. 

The probability of
$\nu_\alpha\to\nu_\beta$
transitions is given by
\begin{equation}
P_{\nu_\alpha\to\nu_\beta}
=
\left|
\sum_{k=1}^{3}
U_{{\beta}k} \,
e^{ \displaystyle -i \frac{ \Delta{m}^2_{k1} L }{ 2 p } } \,
U_{{\alpha}k}^*
\right|^2
\,,
\label{031}
\end{equation}
where $p$ is the neutrino momentum and $L$ is the distance between
the neutrino source and detector.
Let us consider SBL neutrino oscillation experiments.
Taking into account that in these experiments
\begin{equation}
\frac{ \Delta{m}^2_{21} L }{ 2 p } \ll 1
\label{032}
\end{equation}
and using the unitarity of the mixing matrix,
we obtain
\begin{equation}
P_{\nu_\alpha\to\nu_\beta}
=
\left|
\delta_{\alpha\beta}
+
U_{\beta3} \, U_{\alpha3}^*
\left(
e^{ \displaystyle -i \frac{ \Delta{m}^2 L }{ 2 p } } - 1
\right)
\right|^2
\,.
\label{033}
\end{equation}
Thus, under the condition (\ref{032}),
the SBL transition probabilities 
are determined only by the largest mass squared difference
$\Delta{m}^2$
and by the elements of the mixing matrix that connect flavour neutrinos 
with the heaviest neutrino $\nu_3$.

From the expression (\ref{033}),
for the probability of
$\nu_\alpha\to\nu_\beta$
transitions with $\beta\neq\alpha$
and for the survival probability of $\nu_\alpha$ we find
\begin{eqnarray}
P_{\nu_\alpha\to\nu_\beta}
\null & \null = \null & \null
\frac{1}{2}
\,
A_{\beta;\alpha}
\left( 1 - \cos\frac{ \Delta{m}^2 L }{ 2 p } \right)
\,,
\qquad \mbox{for} \qquad
\beta\neq\alpha
\,,
\label{08}
\\
P_{\nu_\alpha\to\nu_\alpha}
\null & \null = \null & \null
1
-
\frac{1}{2}
\,
B_{\alpha;\alpha}
\left( 1 - \cos\frac{ \Delta{m}^2 L }{ 2 p } \right)
\label{09}
\end{eqnarray}
with the oscillation amplitudes
$A_{\beta;\alpha}$
and
$B_{\alpha;\alpha}$
given by
\begin{eqnarray}
A_{\beta;\alpha}
\null & \null = \null & \null
4 \, |U_{\beta3}|^2 \, |U_{\alpha3}|^2
\,,
\label{10}
\\
B_{\alpha;\alpha}
\null & \null = \null & \null
\sum_{\beta\neq\alpha} A_{\beta;\alpha}
=
4 \, |U_{\alpha3}|^2 \left( 1 - |U_{\alpha3}|^2 \right)
\,.
\label{11}
\end{eqnarray}

In the case of a hierarchy of neutrino masses,
neutrino oscillations in SBL experiments
are characterized by
only one oscillation length.
It is obvious that
the dependence of the transition probabilities on the quantity
$ \Delta{m}^2 L / 2 p $
has the same form as in the standard two-neutrino case.
Let us stress, however, that the expressions (\ref{08}) and (\ref{09})
describe transitions between all three flavour neutrinos.
Notice also that in the case of a hierarchy of neutrino masses
the CP phase does not enter in the expressions for the transition 
probabilities.
As a result we have
\begin{equation}
P_{\nu_\alpha\to\nu_\beta}
=
P_{\bar\nu_\alpha\to\bar\nu_\beta}
\label{12}
\end{equation}
in SBL experiments.
As it is seen from Eqs.(\ref{08})--(\ref{11}),
in the scheme under consideration
the oscillations in all channels
($ \nu_e \leftrightarrows \nu_\mu $,
$ \nu_\mu \leftrightarrows \nu_\tau $,
$ \nu_e \leftrightarrows \nu_\tau $)
are described by three parameters:
$\Delta{m}^2$,
$|U_{e3}|^2$,
$|U_{\mu3}|^2$
(because of unitarity of 
the mixing matrix
$ |U_{\tau3}|^2 = 1 - |U_{e3}|^2 - |U_{\mu3}|^2 $).

With the help of Eqs.(\ref{09}) and (\ref{11}),
one can obtain bounds
on the mixing parameters
$|U_{e3}|^2$ and $|U_{\mu3}|^2$
from exclusive plots that were found from the data
of SBL reactor and accelerator disappearance experiments.

We will consider the range
\begin{equation}
10^{-1} \, \mathrm{eV}^2
\leq
\Delta{m}^2
\leq
10^{3} \, \mathrm{eV}^2
\,.
\label{13}
\end{equation}
From the exclusion curves
of SBL disappearance experiments,
at any fixed value of
$\Delta{m}^2$
we obtain the upper bounds 
$ B_{\alpha;\alpha} \leq B_{\alpha;\alpha}^0 $
for $\alpha=e,\mu$.
From Eq.(\ref{11}),
for the mixing parameters
$|U_{\alpha3}|^2$
we have
\begin{equation}
|U_{\alpha3}|^2 \leq a_{\alpha}^0
\qquad \mbox{or} \qquad
|U_{\alpha3}|^2 \geq 1 - a_{\alpha}^0
\,,
\qquad \mbox{with} \qquad
a_{\alpha}^0
=
\frac{1}{2}
\left( 1 - \sqrt{ 1 - B_{\alpha;\alpha}^0 } \,\right)
\,.
\label{14}
\end{equation}
We have obtained the values of $a_{e}^0$ and $a_{\mu}^0$,
respectively,
from the exclusion plots of 
the Bugey reactor experiment \cite{Bugey95}
and the CDHS \cite{CDHS84} and CCFR \cite{CCFR84}
accelerator experiments
(see Fig.1 of \cite{BBGK96}).
In the  range (\ref{13})  
we have
$ a_{e}^0 \lesssim 4 \times10^{-2} $
and
$ a_{\mu}^0 \lesssim 2 \times 10^{-1} $
(for $ \Delta{m}^2 \gtrsim 0.3 \, \mathrm{eV}^2 $).
Thus, from the results of disappearance experiments
it follows that the mixing parameters
$|U_{e3}|^2$
and
$|U_{\mu3}|^2$
can be either small or large (close to one).

Now let us take into account the results of solar neutrino 
experiments.
The probability of solar neutrinos to survive in the case
of a neutrino mass hierarchy is given by \cite{SS92}
\begin{equation}
P_{\nu_e\to\nu_e}^{\mathrm{sun}}(E)
=
\left(
1
-
|U_{e3}|^2
\right)^2
P_{\nu_e\to\nu_e}^{(1,2)}(E)
+
|U_{e3}|^4
\,,
\label{16}
\end{equation}
where
$E$ is the neutrino energy
and
$ P_{\nu_e\to\nu_e}^{(1,2)}(E) $
is the two-generation survival probability of solar $\nu_{e}$'s.
If
$ |U_{e3}|^2 \geq 1 - a_{e}^0 $,
from (\ref{16})
it follows that at all solar neutrino energies
$ P_{\nu_e\to\nu_e}^{\mathrm{sun}} \gtrsim 0.92 $.
This is not compatible with the results of solar neutrino
experiments.
Thus, the mixing parameter
$|U_{e3}|^2$
must be small:
$ |U_{e3}|^2 \leq a_{e}^0 $.

We come to the conclusion that from the results 
of SBL inclusive experiments and solar neutrino
experiments it follows that in the case of three massive neutrinos
with a hierarchy of masses 
only two schemes are possible:
\begin{equation}
\mathrm{I.}
\left\{
\begin{array}{l} \displaystyle
|U_{e3}|^2 \leq a_{e}^0
\,,
\\ \displaystyle
|U_{\mu3}|^2 \leq a_{\mu}^0
\,,
\end{array}
\right.
\qquad \qquad
\mathrm{II.}
\left\{
\begin{array}{l} \displaystyle
|U_{e3}|^2 \leq a_{e}^0
\,,
\\ \displaystyle
|U_{\mu3}|^2 \geq 1 - a_{\mu}^0
\,.
\end{array}
\right.
\label{17}
\end{equation}

Let us consider
$\nu_\mu\leftrightarrows\nu_e$
oscillations
in the case of scheme I.
From Eqs.(\ref{10}) and (\ref{17}),
for the oscillation amplitude we have
\begin{equation}
A_{e;\mu}
\leq
4 \, |U_{e3}|^2 \, |U_{\mu3}|^2
\leq
4 \, a_{e}^0 \, a_{\mu}^0
\,.
\label{18}
\end{equation}
Thus,
in the case of scheme I
the upper bound for the amplitude $A_{e;\mu}$
is quadratic in the small quantities  
$a_{e}^0$, $a_{\mu}^0$
and
$\nu_\mu\leftrightarrows\nu_e$
oscillations are strongly suppressed.

\begin{table}[t!]
\begin{tabular*}{\textwidth}{@{\extracolsep{\fill}}cc}
\begin{minipage}{0.49\linewidth}
\begin{center}
\mbox{\epsfig{file=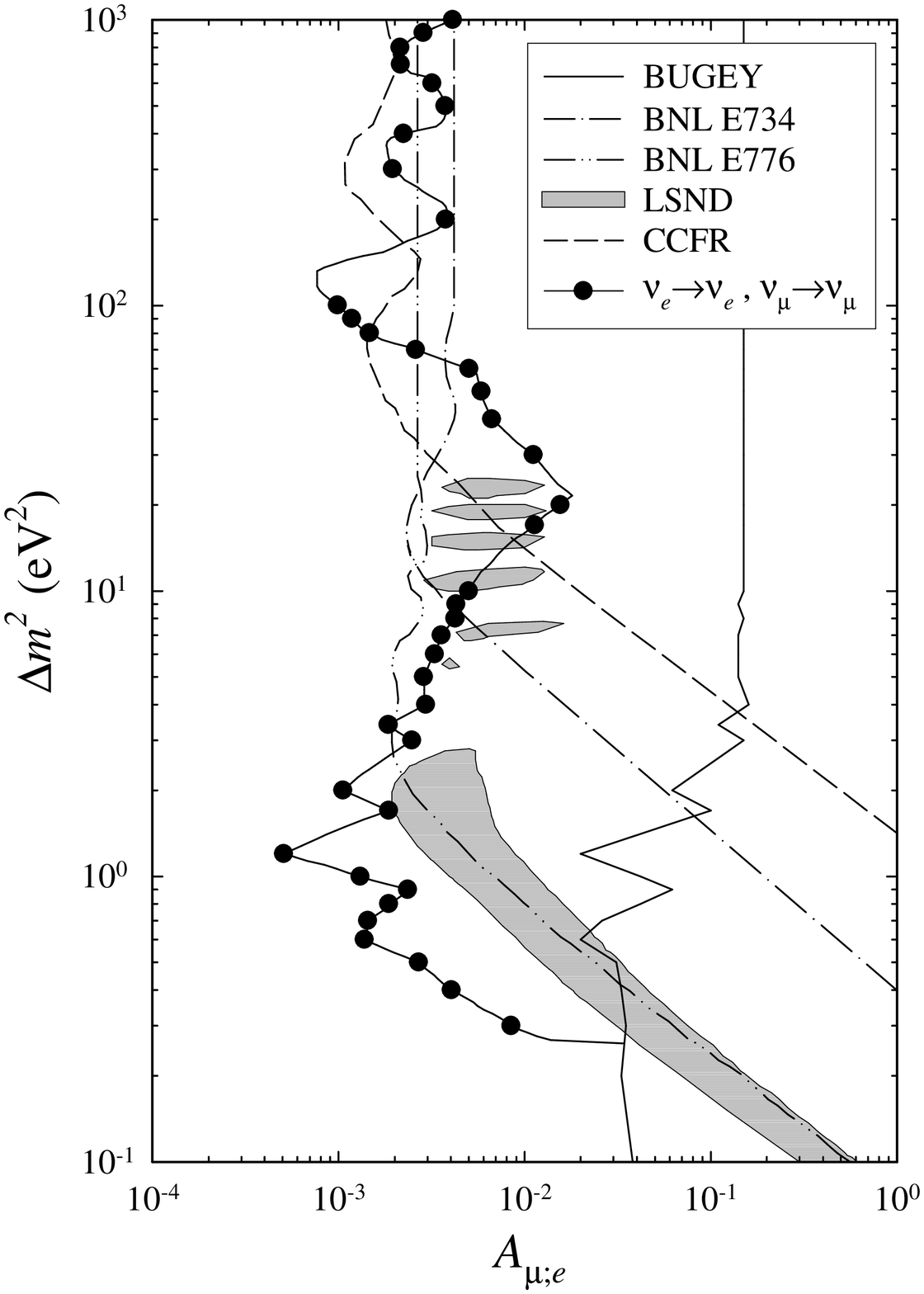,width=0.95\linewidth}}
\end{center}
\end{minipage}
&
\begin{minipage}{0.49\linewidth}
\begin{center}
\mbox{\epsfig{file=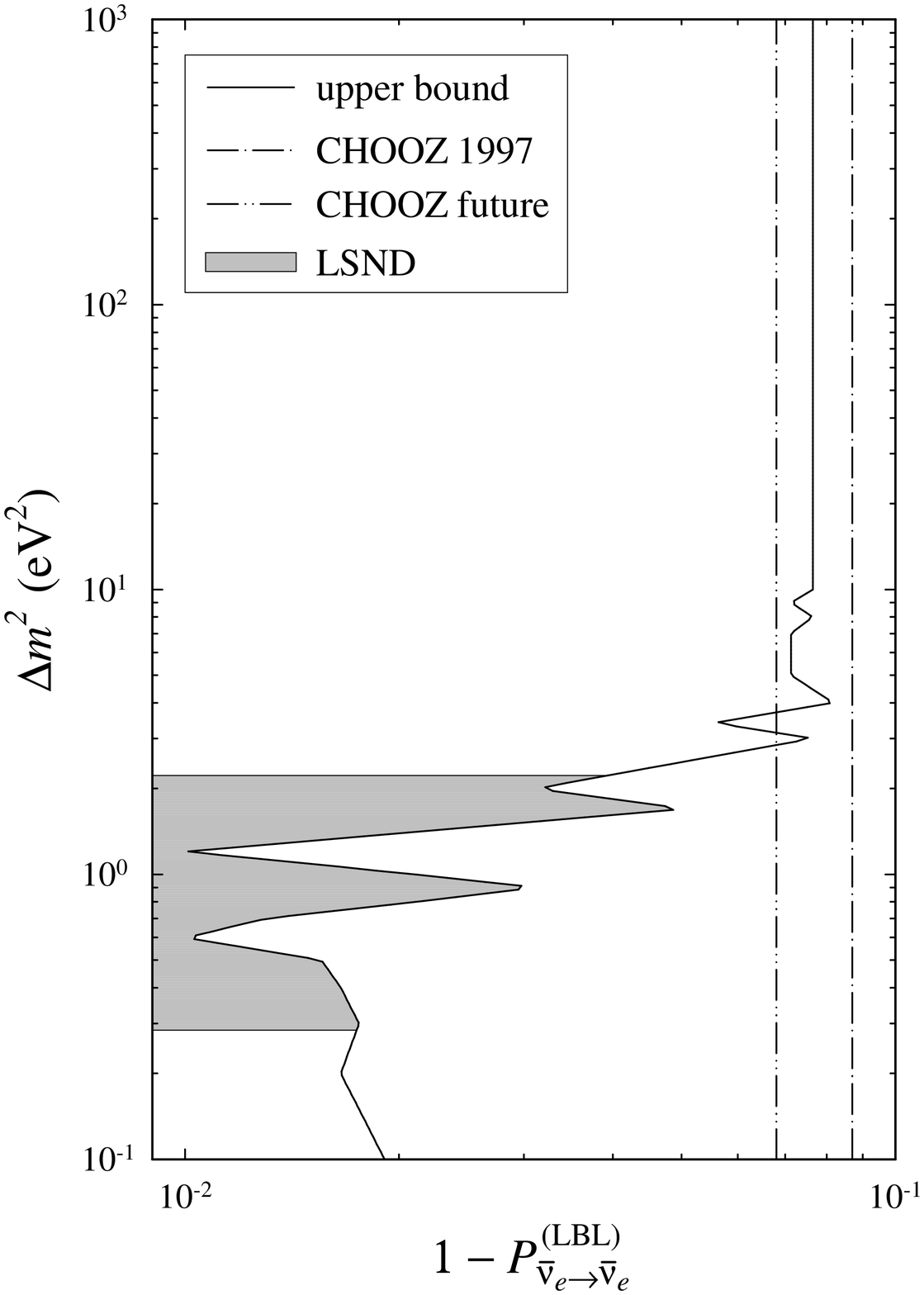,width=0.95\linewidth}}
\end{center}
\end{minipage}
\\
\refstepcounter{figure}
\label{fig1}                 
Figure \ref{fig1}
&
\refstepcounter{figure}
\label{fig2}                 
Figure \ref{fig2}
\end{tabular*}
\null \vspace{-0.5cm} \null
\end{table}

Let us compare now the upper bound (\ref{18})
with the results of the LSND experiment
in which
$\nu_\mu\leftrightarrows\nu_e$
oscillations were observed.
In Fig.\ref{fig1}
the shadowed region in the
$A_{e;\mu}$--$\Delta{m}^2$ plane
is the region allowed at 90 \% CL
by the results of the LSND experiment. 
The regions excluded by the Bugey experiment \cite{Bugey95}
and by the
BNL E734
\cite{BNLE734},
BNL E776
\cite{BNLE776}
and
CCFR
\cite{CCFR96}
experiments are also shown.
The upper bound (\ref{18}) is presented by the curve passing through
the circles.
As it is seen from Fig.\ref{fig1},
the upper bound (\ref{18}) is not
compatible with the results of the LSND experiment if the results of other
oscillation experiments are taken into account.
Thus, the scheme I with a hierarchy of masses and couplings,
similar to the hierarchy that takes place in the quark sector,
is not favoured by the results of SBL experiments.

In the case of scheme II,
the upper bound of the amplitude $A_{e;\mu}$
is linear in the small quantity $a_e^0$:
$ A_{e;\mu} \leq 4 a_e^0 $.
This upper bound is compatible with the LSND data.
Note that, if scheme II is realized in nature,
$ | \nu_\mu \rangle \simeq | \nu_3 \rangle $
and the vectors
$ | \nu_e \rangle $
and
$ | \nu_\tau \rangle $
are superpositions of
$ | \nu_1 \rangle $
and
$ | \nu_2 \rangle $.
For the ``effective'' neutrino masses we have 
$ m_{\nu_\mu} \simeq m_3 $,
$ m_{\nu_e} , m_{\nu_\tau} \ll m_3 $.

Up to now we did not consider atmospheric neutrinos.
In the framework
of the scheme with three massive neutrinos and a neutrino mass 
hierarchy there are only two possibilities to take into account 
the atmospheric neutrino anomaly:
\begin{enumerate}
\item
To assume that
$\Delta{m}^2_{21}$
is relevant for the suppression of solar 
$\nu_e$'s and for the atmospheric neutrino anomaly
\cite{AP97,FLMS97}.
\item
To assume that
$\Delta{m}^2_{31}$
is relevant for the LSND anomaly and 
for the atmospheric neutrino anomaly \cite{CF97,FLMS97}.
\end{enumerate}
The first case is excluded by the results
of the CHOOZ experiment,
that rule out large atmospheric $\nu_\mu\leftrightarrows\nu_e$
transitions.
Other indications against the first case are:
a) the average survival probability of solar $\nu_e$'s is constant 
(this is disfavoured
by the data of
solar neutrino experiments \cite{KP97,CMMV97})
and
b) the parameters
$|U_{e3}|^2$, $|U_{\mu3}|^2$
satisfy the inequalities
$ |U_{e3}|^2 \leq a_e^0 $ and $ |U_{\mu3}|^2 \leq a_\mu^0 $
(that are not compatible with the LSND result, as we have discussed above).

In the second case
it is not possible to explain
the angular dependence of the double ratio of muon and electron events
that was observed by the Kamiokande and Super-Kamiokande experiments 
\cite{Kam-atm,SK}.

All the existing indications in favour of neutrino mixing will be
checked by several experiments that now are under preparation.
If for the time being we accept them,
we come to the necessity of consideration of schemes
with four massive neutrinos, that include the three flavour  
neutrinos $\nu_e$, $\nu_\mu$, $\nu_\tau$
and a sterile neutrino 
\cite{four,BGKP,BGG96,BGG97a,BGG97b,BGG98}.

\section{Four massive neutrinos}
\label{Four massive neutrinos}

There are six possible types of mass spectra with four neutrinos
that can accommodate 
three different scales of $\Delta{m}^2$.
Let us start with the case of a hierarchy of neutrino masses
$ m_1 \ll m_2 \ll m_3 \ll m_4 $,
assuming that
$\Delta{m}^2_{21}$
is relevant for the suppression
of solar $\nu_e$'s,
$\Delta{m}^2_{31}$
is relevant for the atmospheric neutrino anomaly
and
$\Delta{m}^2_{41}$ is relevant for the oscillations
observed in the LSND experiment.
The SBL transition probabilities are given in this case by the expressions
(\ref{08})--(\ref{11})
with the change
$ |U_{\alpha3}|^2 \to |U_{\alpha4}|^2 $
and
$ \Delta{m}^2 \equiv \Delta{m}^2_{41} \equiv m_4^2 - m_1^2 $.
From SBL inclusive data in the range (\ref{13}) of
$\Delta{m}^2$,
we have
\begin{equation}
|U_{\alpha4}|^2 \leq a_{\alpha}^0
\qquad \mbox{or} \qquad
|U_{\alpha4}|^2 \geq 1 - a_{\alpha}^0
\,,
\qquad \mbox{for} \qquad
\alpha=e,\mu
\,,
\label{19}
\end{equation}
with $a_\alpha^0$ given by Eq.(\ref{14}).

For the survival probability of the atmospheric $\nu_\mu$'s
in the scheme under consideration we have the lower 
bound \cite{BGG96}
\begin{equation}
P_{\nu_\mu\to\nu_\mu}^{\mathrm{atm}} \geq |U_{\mu4}|^4
\,.
\label{20}
\end{equation}
Now, from Eq.(\ref{16}) with
$ |U_{e3}|^2 \to |U_{e4}|^2 $
and from Eq.(\ref{20}) it follows that
large values of the mixing parameters
$|U_{e4}|^2$ and $|U_{\mu4}|^2$
are not compatible with solar and atmospheric neutrino data.
We come to the conclusion that both mixing 
parameters
$|U_{e4}|^2$ and $|U_{\mu4}|^2$
are small:
$ |U_{e4}|^2 \leq a_e^0 $ and $ |U_{\mu4}|^2 \leq a_\mu^0 $.
As in the case of scheme I for three neutrinos,
in the scheme under consideration
the SBL amplitude
$A_{e;\mu}$
is constrained by the upper bound (\ref{18})
(with $ |U_{\alpha3}|^2 \to |U_{\alpha4}|^2 $),
which is not compatible with the LSND result
(see Fig.\ref{fig1}). 
Thus, 
a mass hierarchy of four neutrinos is not favoured by the existing
data. The same conclusion can be drawn for all four-neutrino mass
spectra with one neutrino mass separated from the group of three
close masses by the ``LSND gap'' ($\sim 1$ eV). 
 
Let us consider now the two remaining neutrino mass spectra
\begin{equation}
(\mathrm{A})
\qquad
\underbrace{
\overbrace{m_1 < m_2}^{\mathrm{atm}}
\ll
\overbrace{m_3 < m_4}^{\mathrm{sun}}
}_{\mathrm{LSND}}
\,,
\qquad \qquad
(\mathrm{B})
\qquad
\underbrace{
\overbrace{m_1 < m_2}^{\mathrm{sun}}
\ll
\overbrace{m_3 < m_4}^{\mathrm{atm}}
}_{\mathrm{LSND}}
\,,
\label{21}
\end{equation}
with two groups of close masses separated by a $\sim 1$ eV gap.
In the case of such neutrino mass spectra,
the SBL transition probabilities are given by the expressions
(\ref{08}) and (\ref{09})
and the
oscillation amplitudes are given by
\begin{equation}
A_{\beta;\alpha}
=
4 \left| \sum_i U_{{\beta}i} \, U_{{\alpha}i}^{*} \right|^2
\,,
\qquad
B_{\alpha;\alpha}
=
4
\left( \sum_i |U_{{\alpha}i}|^2 \right)
\left( 1 - \sum_i |U_{{\alpha}i}|^2 \right)
\,.
\label{22}
\end{equation}
where the index $i$ runs over $1,2$ or $3,4$.
From the exclusion plots of the
$\bar\nu_e$ and $\nu_\mu$
disappearance experiments we have
($i=1,2$ or $i=3,4$)
\begin{equation}
\sum_i |U_{{\alpha}i}|^2 \leq a_{\alpha}^0
\qquad \mbox{or} \qquad
\sum_i |U_{{\alpha}i}|^2 \geq 1 - a_{\alpha}^0
\,,
\qquad \mbox{for} \qquad
\alpha=e,\mu
\,.
\label{23}
\end{equation}

In the scheme A,
for the survival probabilities
of solar $\nu_e$'s
and atmospheric $\nu_\mu$'s
we have the lower bounds
\cite{BGG96}
\begin{equation}
P_{\nu_e\to\nu_e}^{\mathrm{sun}}
\geq
\left( \sum_{i=1,2} |U_{ei}|^2 \right)^2
\,,
\qquad
P_{\nu_\mu\to\nu_\mu}^{\mathrm{atm}}
\geq
\left( \sum_{i=3,4} |U_{{\mu}i}|^2 \right)^2
\,.
\label{231}
\end{equation}
Hence, the results of solar and atmospheric 
neutrino experiments exclude large values of
$ \displaystyle \left( \sum_{i=1,2} |U_{ei}|^2 \right)^2 $
and
$ \displaystyle \left( \sum_{i=3,4} |U_{{\mu}i}|^2 \right)^2 $
in the case of scheme A
and only two of the four possibilities in Eq.(\ref{23})
are allowed:
\begin{equation}
\sum_{i=1,2} |U_{ei}|^2 \leq a_e^0
\qquad \mbox{and} \qquad
\sum_{i=3,4} |U_{{\mu}i}|^2 \leq a_\mu^0
\,.
\label{24}
\end{equation}
The corresponding inequalities in the scheme B can be obtained
from Eq.(\ref{24}) with the change
$ 1,2 \leftrightarrows 3,4 $.

Now, for the amplitude of
$\nu_\mu\leftrightarrows\nu_e$
oscillations, 
from Eqs.(\ref{22}) and (\ref{24}),
in both schemes we have the upper bound
\begin{equation}
A_{e;\mu}
=
4 \left| \sum_i U_{{\mu}i} \, U_{ei}^{*} \right|^2
\leq
4
\left( \sum_i |U_{{\mu}i}|^2 \right)
\left( \sum_i |U_{ei}|^2 \right)
\leq
4 \, \mathrm{Min}[ a_e^0 , a_\mu^0 ]
\,,
\label{25}
\end{equation}
with $i=1,2$ or $i=3,4$.
This upper bound is compatible with the LSND result.
Thus, schemes A and B can accommodate all neutrino oscillation data.

The schemes A and B 
give different predictions 
for the neutrino mass
$m(^3\mathrm{H})$
measured in
tritium $\beta$-decay
experiments and for the effective Majorana mass
$ \displaystyle
\langle{m}\rangle
=
\sum_{i=1}^4 U_{ei}^2 m_i
$
that determines the matrix element of neutrinoless double-beta decay.
In scheme A we have
\begin{equation}
m(^3\mathrm{H}) \simeq m_4
\,.
\label{26}
\end{equation}
In the case of scheme B,
the contribution to the beta-spectrum of the term that includes 
the heaviest masses $ m_3 \simeq m_4 $
is suppressed by the factor
$ \sum_{i=3,4} |U_{ei}|^2 \leq a_e^0 \lesssim 4 \times 10^{-2} $.

For the effective Majorana mass
in neutrinoless double-beta decay,
in the schemes A and B we have
\begin{equation}
(\mathrm{A})
\qquad
|\langle{m}\rangle|
\leq
\sum_{i=3,4} |U_{ei}|^2 m_4
\leq
m_4
\,,
\qquad \qquad
(\mathrm{B})
\qquad
|\langle{m}\rangle|
\leq
a_e^0 m_4
\ll
m_4
\,.
\label{27}
\end{equation}

Thus, if scheme A is realized in nature,
the tritium $\beta$-decay experiments and
the experiments on the search 
for neutrinoless double-beta decay can see
the effect of the ``LSND neutrino mass''.  

Finally,
we will consider neutrino oscillations in long-baseline (LBL) experiments
in the framework of the schemes A and B.
We will show that the data SBL experiments imply
rather strong constrains on the LBL probabilities of
$\bar\nu_e\to\bar\nu_e$
and
$\nu_\mu\to\nu_e$
transitions \cite{BGG97a}.
In the scheme A,
for the probability of LBL
$\nu_\alpha\to\nu_\beta$
transitions we have the following expression:
\begin{equation}
P^{(\mathrm{LBL,A})}_{\nu_\alpha\to\nu_\beta}
=
\left|
U_{\beta1}
\,
U_{\alpha1}^{*}
+
U_{\beta2}
\,
U_{\alpha2}^{*}
\,
\exp\!\left(
- i
\frac{ \Delta{m}^{2}_{21} \, L }{ 2 \, p }
\right)
\right|^2
+
\left|
\sum_{k=3,4}
U_{{\beta}k}
\,
U_{{\alpha}k}^{*}
\right|^2
\,.
\label{28}
\end{equation}
The probability of the
$\nu_\alpha\to\nu_\beta$
transitions in scheme B
can be obtained from Eq.(\ref{28}) with the change
$ 1,2 \leftrightarrows 3,4 $.
Let us notice also that the probability of LBL
$\bar\nu_\alpha\to\bar\nu_\beta$
transitions can be obtained from Eq.(\ref{28}) with the change
$ U_{{\alpha}k} \to U_{{\alpha}k}^{*} $.

We consider first neutrino oscillations
in reactor experiments
(CHOOZ \cite{CHOOZ},
Palo Verde \cite{PaloVerde},
Kam-Land \cite{Kam-Land}).
From Eq.(\ref{28}), for the 
probability of
$\bar\nu_e\to\bar\nu_e$
transitions in the schemes A and B we have the 
following lower bounds:
\begin{equation}
(\mathrm{A})
\qquad
P^{(\mathrm{LBL,A})}_{\bar\nu_e\to\bar\nu_e}
\geq
\left( \sum_{i=3,4} |U_{ei}|^2 \right)^2
\,,
\qquad \qquad
(\mathrm{B})
\qquad
P^{(\mathrm{LBL,B})}_{\bar\nu_e\to\bar\nu_e}
\geq
\left( \sum_{i=1,2} |U_{ei}|^2 \right)^2
\,.
\label{29}
\end{equation}
Now, taking into account the unitarity of the mixing matrix, 
we can conclude that the quantities
in the right-hand sides of the two inequalities (\ref{29})
are large.
Indeed, from Eq.(\ref{29}),
for both schemes we have
\begin{equation}
P^{(\mathrm{LBL})}_{\bar\nu_e\to\bar\nu_e}
\geq
\left( 1 - a_e^0 \right)^2
\,.
\label{30}
\end{equation}
For the transition probability of $\bar\nu_e$
into any other state, Eq.(\ref{30}) gives
the upper bound
\begin{equation}
1 - P^{(\mathrm{LBL})}_{\bar\nu_e\to\bar\nu_e}
=
\sum_{\alpha\neq{e}}
P^{(\mathrm{LBL})}_{\bar\nu_e\to\bar\nu_\alpha}
\leq
a_e^0 \left( 2 - a_e^0 \right)
\,.
\label{31}
\end{equation}
The value of $a_e^0$ depends on the SBL parameter
$\Delta{m}^2$.
In Fig.\ref{fig2} we have drawn
the curve corresponding to the upper bound (\ref{31})
for $\Delta{m}^2$
in the interval (\ref{13}). 
The shadowed region in Fig.\ref{fig2} is the region that is allowed
(at 90\% CL)
by the data of the LSND experiment and of the other SBL experiments.
Thus, as it is seen from Fig.\ref{fig2},
in the framework of the schemes A and B,
the existing data put rather severe constraints on the LBL
transition probability of $\bar\nu_e$
into any other state.
The results of the first reactor LBL experiment CHOOZ
have been published recently \cite{CHOOZ}.
The upper bound on the
probability
$
\sum_{\alpha\neq{e}}
P^{(\mathrm{LBL})}_{\bar\nu_e\to\bar\nu_\alpha}
$
obtained from the exclusion plot of the CHOOZ experiment
is shown in Fig.\ref{fig2} (dash-dotted line).
One can see that the result of the CHOOZ experiment
agrees with the upper bound obtained from Eq.(\ref{31}).
In Fig.\ref{fig2} we have also drawn the curve corresponding
to the expected final sensitivity of the CHOOZ experiment
(dash-dot-dotted line).
Taking into account the region allowed by the results of LSND and other
SBL experiments,
from Fig.\ref{fig2} one can see that the observation
of neutrino oscillations in the
$\bar\nu_e\to\bar\nu_e$ channel
is extremely difficult.

From the unitarity of the mixing matrix and the CPT-theorem
it follows that the probability of LBL
$\nu_\mu\to\nu_e$
transitions
is also strongly suppressed.
Indeed, we have
\begin{equation}
P^{(\mathrm{LBL})}_{\nu_\mu\to\nu_e}
=
P^{(\mathrm{LBL})}_{\bar\nu_e\to\bar\nu_\mu}
\leq
a_e^0 \left( 2 - a_e^0 \right)
\,.
\label{32}
\end{equation}
Another upper bound on the probability of LBL
$\nu_\mu\to\nu_e$
transitions
can be obtained from Eqs.(\ref{24}) and (\ref{28}).
For both models we have
\begin{equation}
P^{(\mathrm{LBL})}_{\nu_\mu\to\nu_e}
\leq
a_e^0 + \frac{1}{4} \, A_{e;\mu}
\,.
\label{33}
\end{equation}

\begin{table}[t!]
\begin{tabular*}{\textwidth}{@{\extracolsep{\fill}}cc}
\begin{minipage}{0.49\linewidth}
\begin{center}
\mbox{\epsfig{file=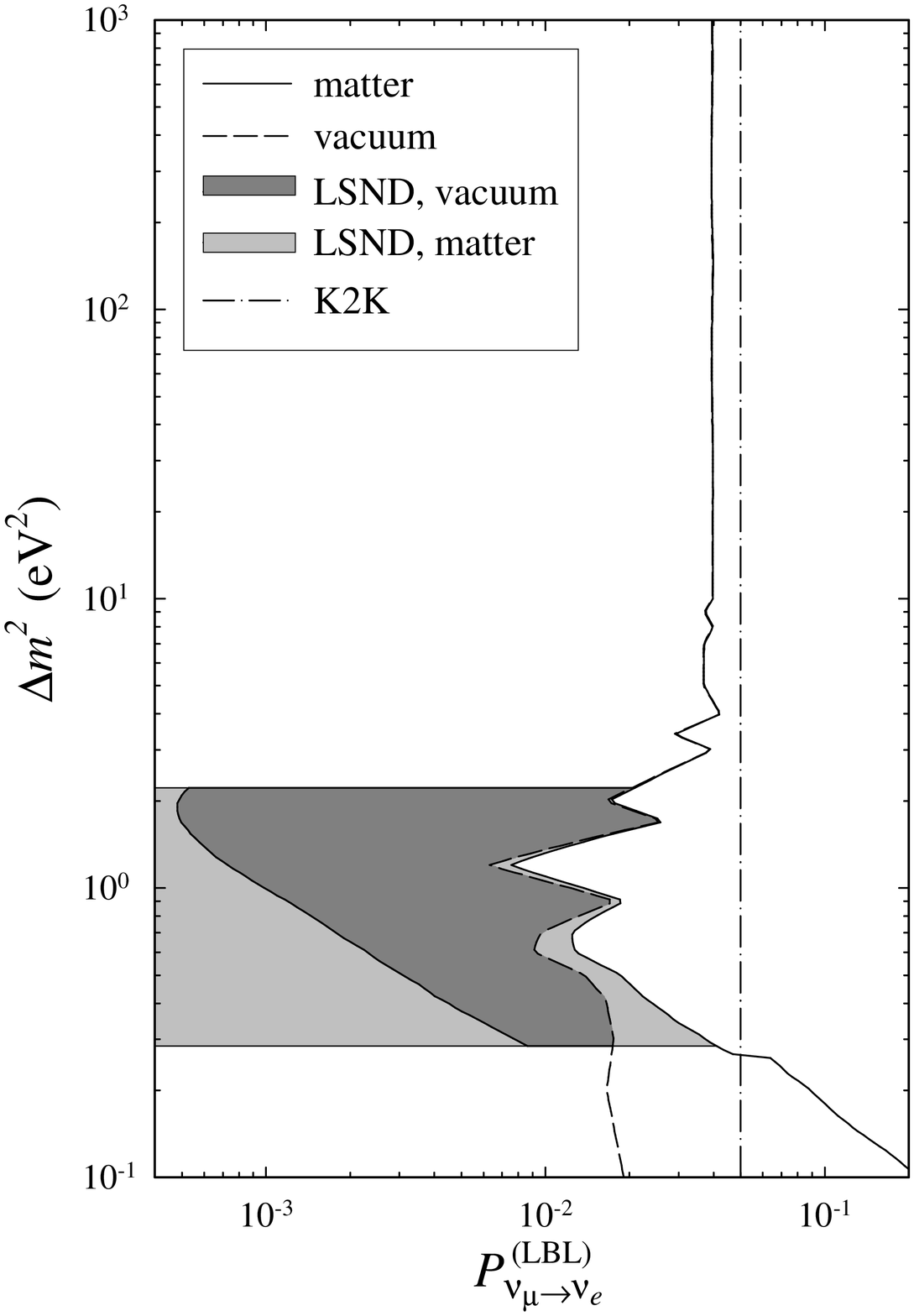,width=0.95\linewidth}}
\end{center}
\end{minipage}
&
\begin{minipage}{0.49\linewidth}
\begin{center}
\mbox{\epsfig{file=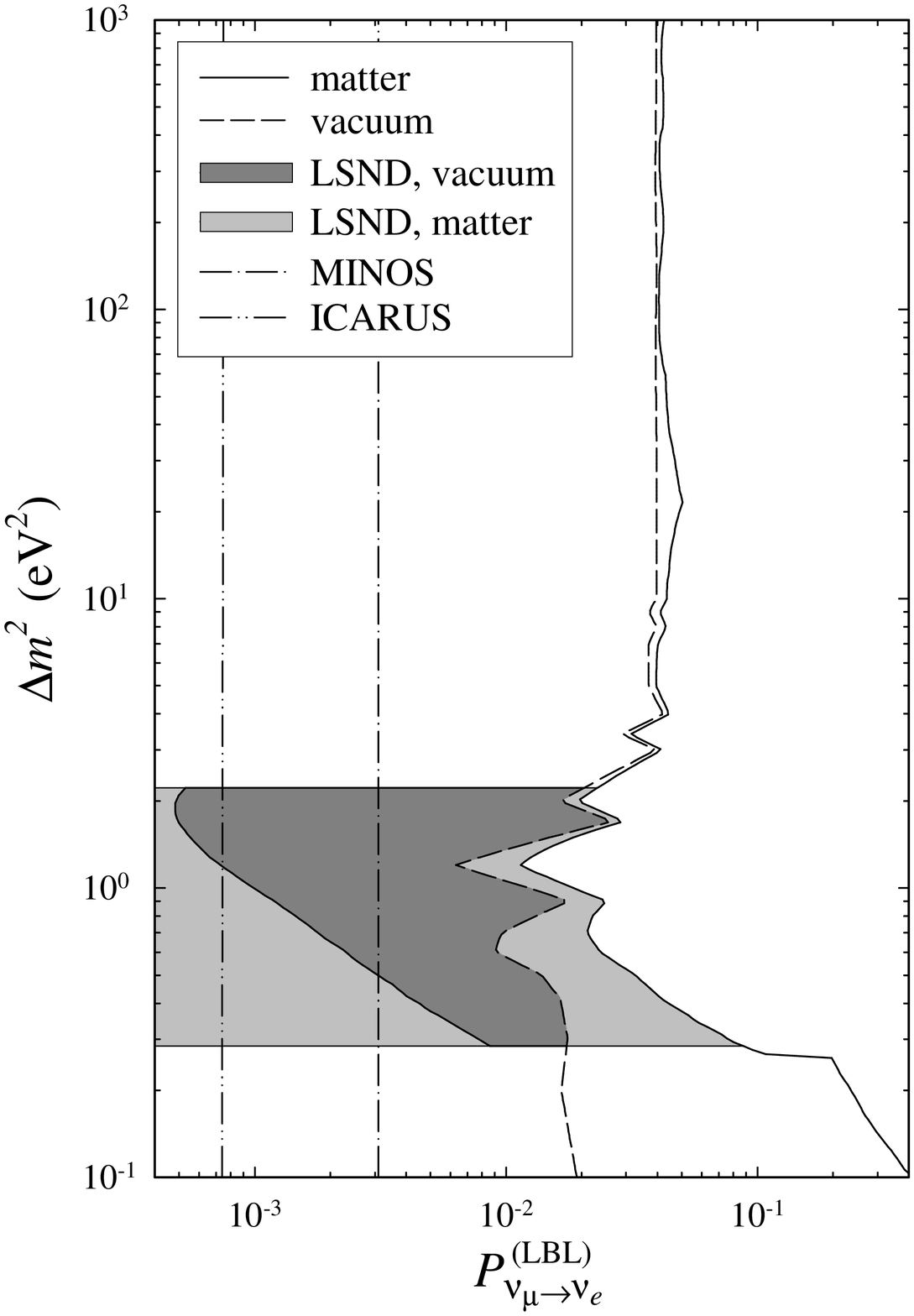,width=0.95\linewidth}}
\end{center}
\end{minipage}
\\
\refstepcounter{figure}
\label{fig3}                 
Figure \ref{fig3}
&
\refstepcounter{figure}
\label{fig4}                 
Figure \ref{fig4}
\end{tabular*}
\null \vspace{-0.5cm} \null
\end{table}

The upper bound for the probability of LBL
$\nu_\mu\to\nu_e$
transitions,
obtained with the help of Eqs.(\ref{32}) and (\ref{33}),
is shown in Fig.\ref{fig3} by the short-dashed curve.
The solid line represents
the corresponding bound with matter corrections
for the K2K experiment \cite{K2K}.
The dash-dotted vertical line 
represents the minimal value of the probability
$P^{(\mathrm{LBL})}_{\nu_\mu\to\nu_e}$
that is expected to be reached in the sensitivity of the K2K experiment.
Notice that at all the considered values of
$\Delta{m}^2$
this probability is larger 
than the upper bound with matter corrections.
The shadowed region in Fig.\ref{fig3}
is allowed at 90\% CL
by the results of LSND and other SBL experiments.
The solid line in Fig.\ref{fig4} shows the bound corresponding
to Eqs.(\ref{32}) and (\ref{33})
with matter corrections
for the MINOS \cite{MINOS} and ICARUS \cite{ICARUS} experiments,
whose expected sensitivities are represented, respectively, by the
dash-dotted and dash-dot-dotted lines.
One can see that these sensitivities are sufficient to explore
the shadowed region allowed by the results of LSND and other SBL
experiments.

\section{Conclusions}
\label{Conclusions}

In the last years there was a big progress in the investigation of the 
problem of neutrino mixing.
Different indications in 
favour of nonzero neutrino masses and mixing angles
have been found. 
The important problem for the experiments of the next 
generation is a detailed investigation of neutrino oscillations 
especially in the 
regions of $\Delta{m}^2$
in which at present there are indications in
favour of oscillations.
Many neutrino experiment are
taking data, or going to start,
or are under preparation:
solar neutrino experiments
(SNO,
ICARUS,
Borexino,
GNO
and others
\cite{future-sun}),
LBL reactor
(CHOOZ \cite{CHOOZ},
Palo Verde \cite{PaloVerde},
Kam-Land \cite{Kam-Land})
and accelerator
(K2K \cite{K2K}, MINOS \cite{MINOS}, ICARUS \cite{ICARUS}
and others \cite{otherLBL})
experiments,
SBL experiments
(CHORUS \cite{CHORUS},
NOMAD \cite{NOMAD},
LSND \cite{LSND},
KARMEN \cite{KARMEN},
BooNE \cite{BooNE})
and many others.
Hence, we have reasons to 
believe that in a few years we will know much more than now about
the fundamental properties of neutrinos
(masses, mixing, their nature (Dirac or Majorana?), etc.).


\begin{thebibliography}{99}

\bibitem{BP78}
S.M. Bilenky and B. Pontecorvo,
Phys. Rep. \textbf{41}, 225 (1978).

\bibitem{BP87}
S.M. Bilenky and S.T. Petcov,
Rev. Mod. Phys. \textbf{59}, 671 (1987).

\bibitem{Mohapatra-Pal}
R.N. Mohapatra and P.B. Pal,
\textit{Massive Neutrinos in Physics and Astrophysics},
World Scientific Lecture Notes in Physics, Vol. 41
(Singapore, 1991).

\bibitem{CWKim}
C.W. Kim and A. Pevsner,
\textit{Neutrinos in Physics and Astrophysics},
Contemporary Concepts in Physics, Vol.8
(Harwood Academic Press, Chur, Switzerland, 1993).

\bibitem{Homestake}
B.T. Cleveland \textit{et al.},
Nucl. Phys. B (Proc. Suppl.) \textbf{38}, 47 (1995).

\bibitem{Kam-sun}
K.S. Hirata \textit{et al.},
Phys. Rev. D \textbf{44}, 2241 (1991).

\bibitem{GALLEX}
GALLEX Coll.,
W. Hampel \textit{et al.},
Phys. Lett. B \textbf{388}, 384 (1996).

\bibitem{SAGE}
J.N. Abdurashitov \textit{et al.},
Phys. Rev. Lett. \textbf{77}, 4708 (1996).

\bibitem{SK-sun}
K. Inoue,
Talk presented at the
$5^{\mathrm{th}}$
International Workshop on
\textit{Topics in Astroparticle and Underground Physics},
Gran Sasso, Italy, September 1997 
(http://\-www-sk.\-icrr.\-u-tokyo.\-ac.\-jp/\-doc/\-sk/\-pub/\-pub\_sk.\-html);
R. Svoboda,
Talk presented at the Conference on
\textit{Solar Neutrinos: News About SNUs},
2--6 December 1997, Santa Barbara, California
(http://\-www.\-itp.\-ucsb.\-edu/\-online/\-snu/).

\bibitem{SK}
M. Nakahata,
Talk presented at the
APCTP Workshop:
\textit{Pacific Particle Physics Phenomenology},
Seoul, Korea, October 31 -- November 2, 1997
(http://\-www-sk.\-icrr.\-u-tokyo.\-ac.\-jp/\-doc/\-sk/\-pub/\-pub\_sk.\-html).

\bibitem{Kam-atm}
Y. Fukuda \textit{et al.},
Phys. Lett. B \textbf{335}, 237 (1994).

\bibitem{IMB}
R. Becker-Szendy \textit{et al.},
Nucl. Phys. B (Proc. Suppl.) \textbf{38}, 331 (1995).

\bibitem{Soudan}
W.W.M. Allison \textit{et al.},
Phys. Lett. B \textbf{391}, 491 (1997).

\bibitem{SK-atm}
Super-Kamiokande Coll.,
Y. Fukuda \textit{et al.},
ICRR-Report-411-98-7 (hep-ex/9803006).

\bibitem{LSND}
C. Athanassopoulos \textit{et al.},
Phys. Rev. Lett. \textbf{77}, 3082 (1996).

\bibitem{PDG}
Particle Data Group,
R.M. Barnett \textit{et al.},
Phys. Rev. D \textbf{54}, 1 (1996).

\bibitem{Pontecorvo57}
B. Pontecorvo,
J. Exptl. Theoret. Phys. \textbf{33}, 549 (1957)
[Sov. Phys. JETP \textbf{6}, 429 (1958)];
J. Exptl. Theoret. Phys. \textbf{34}, 247 (1958)
[Sov. Phys. JETP \textbf{7}, 172 (1958)].

\bibitem{see-saw}
M. Gell-Mann, P. Ramond, and R. Slansky,
in \textit{Supergravity},
ed. F. van Nieuwenhuizen and D. Freedman
(North Holland, Amsterdam, 1979), p.315;
T. Yanagida,
Proc. of the
\textit{Workshop on Unified Theory and the Baryon Number of the Universe},
KEK, Japan, 1979;
R.N. Mohapatra and G. Senjanovi\'c,
Phys. Rev. Lett. \textbf{44}, 912 (1980).

\bibitem{HL97}
N. Hata and P. Langacker,
Phys. Rev. D \textbf{56}, 6107 (1997).

\bibitem{FLM97}
G.L. Fogli, E. Lisi and D. Montanino,
preprint hep-ph/9709473.

\bibitem{Valencia}
M.C. Gonzalez-Garcia \textit{et al.},
preprint hep-ph/9801368.

\bibitem{MSW}
S.P. Mikheyev and A.Yu. Smirnov,
Yad. Fiz. \textbf{42}, 1441 (1985)
[Sov. J. Nucl. Phys. \textbf{42}, 913 (1985)];
Il Nuovo Cimento C \textbf{9}, 17 (1986);
L. Wolfenstein,
Phys. Rev. D \textbf{17}, 2369 (1978);
\textit{ibid.} \textbf{20}, 2634 (1979).

\bibitem{Brunner}
J. Brunner,
Fortsch. Phys. \textbf{45}, 343 (1997).

\bibitem{CHOOZ}
M. Apollonio \textit{et al.},
Phys. Lett. B \textbf{420}, 397 (1998).

\bibitem{Troitsk}
A.I. Belesev et al.,
Phys. Lett. B \textbf{350}, 263 (1995);
V.M. Lobashev,
Talk presented at the
\textit{Erice School on Nuclear Physics, $19^{\mathrm{th}}$ course
"Neutrinos in Astro, Particle and Nuclear Physics"},
16--24 September 1997.

\bibitem{Mainz}
C. Weinheimer et al.,
Phys. Lett. B \textbf{300}, 210 (1993).

\bibitem{Heidelberg-Moscow}
M. G\"unther et al.,
Phys. Rev. D \textbf{55}, 54 (1997);
Phys. Lett. B \textbf{407}, 219 (1997).

\bibitem{futureBB}
NEMO Collaboration,
Nucl. Phys. B (Proc. Suppl.) \textbf{48}, 226 (1996);
F.A. Danevich et al.,
\textit{ibid.}, 232 (1996);
A. Alessandrello et al.,
\textit{ibid.}, 238 (1996);
H.V. Klapdor-Kleingrothaus, J. Hellming and M. Hirsch,
J. Phys. G \textbf{24}, 483 (1998).

\bibitem{BBGK95}
S.M. Bilenky, A. Bottino, C. Giunti and C.W. Kim,
Phys. Lett. B \textbf{356}, 273 (1995).

\bibitem{BBGK96}
S.M. Bilenky, A. Bottino, C. Giunti and C.W. Kim,
Phys. Rev. D \textbf{54}, 1881 (1996).

\bibitem{BGKP}
S.M. Bilenky, C. Giunti, C.W. Kim and S.T. Petcov,
Phys. Rev. D \textbf{54}, 4432 (1996).

\bibitem{BGG96}
S.M. Bilenky, C. Giunti and W. Grimus,
Proc. of
\textit{Neutrino96}, Helsinki, June 1996, edited by K. Enqvist
\textit{et al.}, p.174 (World Scientific, Singapore, 1997);
Eur. Phys. J. C \textbf{1}, 247 (1998).

\bibitem{BGG97a}
S.M. Bilenky, C. Giunti and W. Grimus,
Phys. Rev. D \textbf{57} (1998) 1920.

\bibitem{BGKM}
S.M. Bilenky, C. Giunti, C.W. Kim and M. Monteno,
preprint hep-ph/9711400, to be published in Phys. Rev. D.

\bibitem{BGG97b}
S.M. Bilenky, C. Giunti and W. Grimus, 
preprint hep-ph/9712537, to be published in Phys. Rev. D.

\bibitem{BGG98}
S.M. Bilenky, C. Giunti, W. Grimus and T. Schwetz, 
preprint hep-ph/9804421.

\bibitem{three}
A. De Rujula \emph{et al.},
Nucl. Phys. B \textbf{168}, 54 (1980);
V. Barger and K. Whisnant,
Phys. Lett. B \textbf{209}, 365 (1988);
S.M. Bilenky \emph{et al.},
\emph{ibid.} \textbf{276}, 223 (1992);
K.S. Babu \emph{et al.},
\emph{ibid.} \textbf{359}, 351 (1995).
H. Minakata,
\emph{ibid.} \textbf{356}, 61 (1995);
Phys. Rev. D \textbf{52}, 6630 (1995);
G.L. Fogli \emph{et al.},
\emph{ibid.} \textbf{52}, 5334 (1995);
S. Goswami \emph{et al.},
Int. J. Mod. Phys. A \textbf{12}, 781 (1997).

\bibitem{Bugey95}
B. Achkar \emph{et al.},
Nucl. Phys. B \textbf{434}, 503 (1995).

\bibitem{CDHS84}
F. Dydak \emph{et al.},
Phys. Lett. B \textbf{134}, 281 (1984).

\bibitem{CCFR84}
I.E. Stockdale \emph{et al.},
Phys. Rev. Lett. \textbf{52}, 1384 (1984).

\bibitem{SS92}
X. Shi and D.N. Schramm,
Phys. Lett. B \textbf{283}, 305 (1992).

\bibitem{BNLE734}
L.A. Ahrens \emph{et al.},
Phys. Rev. D \textbf{36}, 702 (1987).

\bibitem{BNLE776}
L. Borodovsky \emph{et al.},
Phys. Rev. Lett. \textbf{68}, 274 (1992).

\bibitem{CCFR96}
A. Romosan \emph{et al.},
Phys. Rev. Lett. \textbf{78}, 2912 (1997).

\bibitem{AP97}
A. Acker and S. Pakvasa,
Phys. Lett. B \textbf{357}, 209 (1997).

\bibitem{FLMS97}
G.L. Fogli, E. Lisi, D. Montanino and G. Scioscia,
Phys. Rev. D \textbf{56}, 4365 (1997).

\bibitem{CF97}
C.Y. Cardall and G.M. Fuller,
Phys. Rev. D \textbf{53}, 4421 (1996);
C.Y. Cardall, G.M. Fuller and D.B. Cline,
Phys.Lett. B \textbf{413}, 246 (1997).

\bibitem{KP97}
P.I. Krastev and S.T. Petcov,
Phys. Lett. B \textbf{395}, 69 (1997).

\bibitem{CMMV97}
G. Conforto, A. Marchionni, F. Martelli and F. Vetrano,
preprint hep-ph/9708301.

\bibitem{four}
J.T. Peltoniemi and J.W.F. Valle,
Nucl. Phys. B \textbf{406}, 409 (1993);
D.O. Caldwell and R.N. Mohapatra,
Phys. Rev. D \textbf{48}, 3259 (1993);
Z. Berezhiani and R.N. Mohapatra,
\emph{ibid} \textbf{52}, 6607 (1995);
J.R. Primack et al.,
Phys. Rev. Lett. \textbf{74}, 2160 (1995);
E. Ma and P. Roy,
Phys. Rev. D \textbf{52}, R4780 (1995);
R. Foot and R.R. Volkas,
\emph{ibid} \textbf{52}, 6595 (1995);
E.J. Chun et al.,
Phys. Lett. B \textbf{357}, 608 (1995);
J.J. Gomez-Cadenas and M.C. Gonzalez-Garcia,
Z. Phys. C \textbf{71}, 443 (1996);
E. Ma,
Mod. Phys. Lett. A \textbf{11}, 1893 (1996);
N. Okada and O. Yasuda, 
Int. J. Mod. Phys. A \textbf{12}, 3669 (1997).
S. Goswami,
Phys. Rev. D \textbf{55}, 2931 (1997);
A.Yu. Smirnov and M. Tanimoto,
\emph{ibid} \textbf{55}, 1665 (1997);
V. Barger, T. Weiler and K. Whisnant,
preprint hep-ph/9712495;
E. J. Chun, C. W. Kim and U. W. Lee,
preprint hep-ph/9802209;
S.C. Gibbons, R.N. Mohapatra, S. Nandi and A. Raychaudhuri,
preprint hep-ph/9803299.

\bibitem{PaloVerde}
F. Boehm et al.,
\textit{The Palo Verde experiment},
1996;
http:\-//\-www.\-cco.\-caltech.\-edu/\~{}songhoon/\-Palo-Verde/\-Palo-Verde.\-html.

\bibitem{Kam-Land}
F. Suekane,
preprint TOHOKU-HEP-97-02, June 1997;
M. Nakahata,
Talk presented at the
\textit{International Europhysics Conference on High Energy Physics},
19-26 August 1997, Jerusalem, Israel
(http://\-www.cern.ch/\-hep97/\-abstract/\-tpl.htm).

\bibitem{K2K}
Y. Suzuki,
Talk presented at
\textit{Neutrino 96},
Helsinki, June 1996;
http://\-pnahp.kek.jp/.

\bibitem{MINOS}
MINOS Coll.,
D. Ayres \emph{et al.},
NUMI-L-63, February 1995;
http://\-www.hep.anl.gov/\-NDK/\-HyperText/\-numi.html.

\bibitem{ICARUS}
ICARUS Coll.,
P. Cennini \emph{et al.},
LNGS-94/99-I,
May 1994;
http://\-www.aquila.infn.it/\-icarus/.

\bibitem{future-sun}
Y. Suzuki (SuperKamiokande),
Proc. of the
\textit{Fourth International Solar Neutrino Conference},
Heidelberg, Germany, 8--11 April 1997,
edited by W. Hampel
(Max-Planck-Institut f\"ur Kernphysik, 1997),
p.163;
R. Meijer Drees (SNO),
\emph{ibid.}, p.210;
F. von Feilitzsch (Borexino),
\emph{ibid.}, p.192;
E. Bellotti (GNO),
\emph{ibid.}, p.173;
K. Lande (Homestake Iodine),
\emph{ibid.}, p.228;
C. Tao (HELLAZ),
\emph{ibid.}, p.238;
A.V. Kopylov (Lithium),
\emph{ibid.}, p.263;
Yu.G. Zdesenko (Xenon),
\emph{ibid.}, p.283;
P. Cennini \emph{et al.} (ICARUS),
LNGS-94/99-I,
May 1994;
T.J. Bowels (GaAs),
Proc. of
\textit{Neutrino 96},
Helsinki, Finland, 13--19 June 1996,
edited by K. Enqvist, K. Huitu and J. Maalampi
(World Scientific, Singapore, 1997),
p.83.

\bibitem{otherLBL}
NOE Coll., G.C. Barbarino \emph{et al.},
INFN/AE-96/11, May 1996;
AQUA-RICH Coll., T. Ypsilantis \emph{et al.},
LPC/96-01, CERN-LAA/96-13;
OPERA Coll., H. Shibuya \emph{et al.},
CERN-SPSC-97-24.

\bibitem{CHORUS}
D. Macina,
Nucl. Phys. B (Proc. Suppl.) \textbf{48}, 183 (1996).

\bibitem{NOMAD}
M. Laveder,
Nucl. Phys. B (Proc. Suppl.) \textbf{48}, 188 (1996).

\bibitem{KARMEN}
J. Kleinfeller,
Nucl. Phys. B (Proc. Suppl.) \textbf{48}, 207 (1996).

\bibitem{BooNE}
Booster Neutrino Experiment
(BooNE),
http://\-nu1.lampf.lanl.gov/\-BooNE.

\end{thebibliography}
\end{document}